\documentclass[11pt]{article}
\usepackage[dvips]{graphicx}
\usepackage{float}
\usepackage{setspace}
\usepackage{cite}
\usepackage[all]{xy}

\begin{document}
\doublespacing

\begin{titlepage}

\begin{center}
  {\Large\bf 
  
  Double Periodicity and Frequency-Locking \\
  
  in the Langford Equation\\}

\

  {\Large\bf Short running title: 
  
  Double Periodicity in the Langford Equation\\}

\

\

  {\large Makoto Umeki}
  
  {E-mail address: umeki@phys.s.u-tokyo.ac.jp}
  
  {\large Department of Physics, Graduate School of Science}

  {\large University of Tokyo, 7-3-1 Hongo, Bunkyo-ku, Tokyo 113-0033}

\end{center}

\ \ 

\newpage

\begin{abstract}

\large

The bifurcation structure of the Langford equation is studied numerically in detail. 
Periodic, doubly-periodic, and chaotic solutions and the routes to chaos 
via coexistence of double periodicity and period-doubling bifurcations 
are found by the Poincar\'e plot of successive maxima of 
the first mode $x_1$. Frequency-locked periodic solutions corresponding 
to the Farey sequence $F_n$ are examined up to $n=14$. 
Period-doubling bifurcations appears on some of the periodic solutions 
and the similarity of bifurcation structures between the sine-circle 
map and the Langford equation is shown. 
A method to construct the Poincar\'e section for triple periodicity 
is proposed. 
\end{abstract}


\end{titlepage}

\section{Introduction}

A quasi-periodicity route to turbulence due to Ruelle and Takens \cite{Ruelle}
is that transition could occur via three successive Hopf 
bifurcations, leading from a fixed point to a limit cycle, 
then to a torus and finally to a 3-torus. 
On the other hand, double periodicity is considered to be 
well modeled by the one dimensional sine-circle map, 
which indicates that frequency-locking and periodic solutions 
appear as the nonlinear parameter increases. 
Therefore, it is a natural question what actually happens on attractors of 
a simple set of ordinary differential equations (ODEs) having double periodicity, 
like the Langford equation \cite{Langford1,Langford2,Langford3}. 
Similar numerical studies have been done for the five dimensional 
ODEs modeling magnetoconvection \cite{Bekki1,Bekki2} and the 
six dimensional ODEs of the Gledzer shell model of turbulence \cite{Umeki1}. 
In the latter case the parameter to be changed is the viscosity, or 
equivalently, the Reynolds number. 
Both studies show a bifurcation structure very similar to that of 
the sine-circle map, although triple periodicity 
is stated based on the numerically obtained Lyapunov exponents in \cite{Bekki2}.  
The presented results correspond to a detailed study that extends 
Langford \cite{Langford1,Langford2,Langford3}. 

In Section 2, a method to construct the three dimensional Poincar\'e 
section is proposed. For triple periodicity expected by 
Ruelle and Takens \cite{Ruelle}, points lie on a surface. 
Frequency-locked double periodicity indicates points on a closed curve
embedded on the surface. 
Section 3 gives the Langford equation with an explanation of 
the evolution of the energy and selected parameters. 
In Section 4, numerical results of bifurcation structures of the 
equation are shown. 
It is confirmed that, instead of triple periodicity, 
we have double periodicity, frequency-locking and period-doubling 
bifurcations. 
The structure is very complicated; 
frequency-locking corresponding to the Farey sequence with the index 
up to 14 is confirmed, and 
plural sequences of the period-doubling bifurcations are observed. 
Summary and further possibilities to explore complexity of 
dynamical systems modeling fluid dynamics and other high dimensional 
systems with the presented approach are described in Conclusions. 

\section{Triple periodicity}

As a typical example of triple periodicity, 
we consider the following function which is a sum of three 
sinusoidal functions of $t$; 

\begin{equation}
x(t;a,\omega )= \sin t + \sin \pi t + a \sin \omega t.
\label{eq1} 
\end{equation}

A standard method to construct the Poincar\'e map used in \cite{Bekki1,Umeki1}
is to seek for successive local maxima $x_n$ of $x(t)$. 
Then we let $z_n=x_n +i x_{n+1}$ and $ \theta_n=(2\pi)^{-1} {\rm Arg} (z_n)$ (mod 1). 
For double periodicity, $\theta_n$ obeys the generalized sine-circle map 
\cite{Umeki1}, which gives a curve on the $(\theta_n,\theta_{n+1})$ plane. 
Similarly, we can anticipate a surface in the 
$(\theta_n,\theta_{n+1},\theta_{n+2})$ space.  
To check this analogy, 
we show the Poincar\'e plot of Eq. (\ref{eq1}) in the time interval $0<t<5000$ 
for double periodicity $a=0$ in Figure 1, 
triple periodicity $a=0.5, \omega=\sqrt{2}$ in Figure 2, 
and frequency-locked double periodicity $a=0.5, \omega=1.4$ in Figure 3. 
Points lie on a curve in Figure 1 (a) and on a torus in Figure 2 (a), 
whose three dimensional structure can be viewed by a new 3D graphics function 
in {\it Mathematica} 6.  
The latter indicates a possibility that $\theta_{n+2}$ can be expressed by 
a function of $\theta_n$ and $\theta_{n+1}$. 
This Poincar\'e plot will give a method to identify triple periodicity 
among complicated time series of general numerical or observed data. 

We have 5:7 frequency-locking in the case of Figure 3. Correspondingly, 
points are again on a curve, which is more complicated than that in Figure 1. 
Actual data may contain higher harmonics in Eq. (\ref{eq1}) 
but the qualitative behavior can be expected to be similar. 

\section{The Langford Equation}

The Langford equation is the set of three ordinary differential equations 
for $x_i(t)$, $i=1,2,3$, given as follows:
\begin{eqnarray}
\frac{dx_1}{dt}=& \dot{x}_1 = & (x_3-b) x_1 -c x_2, \label{eq2} \\
\frac{dx_2}{dt}=& \dot{x}_2 = & c x_1 + (x_3-b) x_2, \label{eq3} \\
\nonumber
\frac{dx_3}{dt}=& \dot{x}_3 = & d+ a x_3 - 
\displaystyle 
\frac{x_3^3}{3} \\
& & -(x_1^2+x_2^2)(1+f x_3)
+ e x_3 x_1^3. \label{eq4} 
\end{eqnarray}

The temporal evolution of the {\it energy} defined by $ E=(1/2) \sum_{i=1}^3 x_i^2$ 
is given by

\begin{equation}
\frac{dE}{dt}= d x_3+a x_3^2-\frac{x_3^4}{3} -(b+f x_3^2)(x_1^2+x_2^2)
+ e x_3^2 x_1^3.
\label{eq5} 
\end{equation}

The set of the parameters fixed in this paper is borrowed from \cite{Langford3} as 
\begin{equation}
(a,b,c,d,f)=(1,0.7,3.5,0.6,0.25), \label{eq6}
\end{equation}
and $e$ is the changing parameter. 
Because $b>0$ and $f>0$, the right hand side of (\ref{eq5}) becomes 
negative if $E$ is sufficiently large for $e=0$, leading to the proof 
that the solution is finite. 
If $e\ne 0$, the finiteness of the solution is unknown 
but supported by the numerical solution for small $e$.

The contraction rate of the volume element in the phase space 
is given by
\begin{equation}
\frac{\partial \dot{x}_1}{\partial x_1} +\frac{\partial \dot{x}_2}{\partial x_2} 
+\frac{\partial \dot{x}_3}{\partial x_3} =
-x_3^2+2 x_3-2b+a -f(x_1^2+x_2^2)+e x_1^3.  \label{eq7}
\end{equation}
It depends on the position in the phase space. 
If $e=0$, it becomes negative for sufficiently large values of $E$
since $f$ is positive and the quadratic terms become dominant. 

\section{Bifurcation Structure}

Before we show numerical results of the Langford equation, it is useful to 
review the bifurcation structure of the sine-circle map:
\begin{equation}
\theta_{n+1}=f(\theta_n) = \theta_n+ \Omega + \frac{K}{2\pi} \sin (2\pi \theta_n), 
\qquad ({\rm mod} \ \  1). 
\label{eq8}
\end{equation}
In order to compare the diagram of (\ref{eq8}) with that of the Langford equation, 
the parameter $\Omega$ and $K$ are parametrized as
\begin{equation}
\Omega= 0.5(1+\alpha ), \qquad K=4 \alpha.
\label{eq9}
\end{equation}
The initial condition is $\theta_1=0$, $0<\alpha<1$, $\Delta \alpha=0.001$, 
the total iteration is 400, and the last 200 steps are plotted in Figure 4. 
Although double periodicity and frequency-locking are illustrated by the sine-circle map 
in many literatures, the period-doubling bifurcation is also observed at $\alpha \simeq 0.6$ 
in Figure 4, which can make it easy to understand the similarity of 
the bifurcation structures between the sine-circle map and the Langford equation. 
The coexistence of two scenarios of routes to turbulence, 
quasi periodicity and period-doubling bifurcation, may be common in 
many of dynamical systems. 

In order to study bifurcation structures of the Langford equation, 
sets of $x_1$ at its local maximum after transient states 
are plotted with various values of the parameter $e$. 
Eqs. (\ref{eq2}-\ref{eq4}) are solved by {\it NDsolve} command in {\it Mathematica} 5.2, 
which makes 100 to 500 different computations possible in a single 
program. 
The typical final time is $t_f=2000\sim 4000$ and the last period 
$t_f-t_d \le t\le t_f $ with $t_d=200 \sim 600$ is picked up to identify attractors. 
In order to find the local maximum, the time period is divided into 
intervals with the width $\Delta t=0.1$, and for the 
interval which can include the local maximum, the {\it FindMaximum} command is 
invoked. 
The typical initial condition for the numerical computation is 
$(x_1,x_2,x_3)=(0.01,0,0)$. We also choose the initial condition 
as the final state just before the parameter varied in order 
to examine the hysteresis. In some cases, multiple stable states 
are observed. 
About 40 different runs are performed with various regions 
of $e$ and suitable numerical parameters. 

Figure 5 shows the bifurcation diagram for the parameter range $0\le e \le 0.2$. 
The step of the parameter $e$ is $\Delta e=0.0001$.
The region $e<0.03$ includes doubly periodic solutions. 
Then 3:4 frequency-locking yields the periodic solution. 
The Feigenbaum period-doubling bifurcation occurs at $e\simeq 0.057$. 
There are many periodic windows as well as chaotic solutions 
after the period-doubling bifurcation. 
Its behavior is very similar to that of the one dimensional logistic map. 

The range $0.02<e<0.03$ is enlarged in Figure 6, in order to see 
the most dominant frequency-locked periodic solutions. 
The solution is doubly periodic if $e<0.02028$, even if $e$ is negative. 
As $e$ increases, a periodic solution with 17:23 resonance 
appears at $e\simeq 0.02028$. 
Periodic solutions shown by $m_2$ points are specified by 
the resonance condition $m_1:m_2$ drawn at just below the 
upper frame of Figure 6. 
The region including double periodicity ends with the emergence of 
3:4 resonance at $e\simeq 0.02935$. 
Doubly periodic solutions draw points whose numbers are 
controlled by the time range for plotting and typically 
much larger than the periodic solutions. 

The Farey sequence $F_n$ for integers $n>0$ is the set of irreducible 
rational numbers $a/b$ where $0\le a \le b\le n$ 
and the greatest common divisor of $a$ and $b$ is 1.
The Farey sequence appears in frequency-locking of the sine-circle map. 
The resonance conditions between 17:23 and 
3:4 are shown up to the Farey index $n=14$ in Table 1. 

For the 23:31 resonant periodic solution, an incomplete 
period-doubling bifurcation is observed in Figure 6.  
Figure 7 shows the diagram enlarged for $0.021<e<0.0235$.
Figure 8 shows the diagram for $0.0243<e<0.0253$. 
The period-doubling bifurcations observed on both of the two 
parameter regions of 43:58 resonance. 

The periodicity can be also judged by numerical 
computation of the square of the minimum distance 
\begin{equation}
D= \min_{t>0} \sum_{i=1}^3 (x_i(t)-x_i(0))^2, 
\label{eq10}
\end{equation}
indicating the accuracy of recurrence. 
In Figure 9 for $0.0272<e<0.0283$, the small value of $D$ indicates 
frequency-locked periodicity including resonances corresponding
to $n=15$ (35:47) and $n=16$ (38:51 and 93:125) of $F_n$. 

The width of resonance decreases as the index $n$ of $F_n$ increases, 
that implies scaling laws as noted by \cite{Bekki1}. 
Summary of observed stable periodic windows is given in Table 2. 
Figure 10 shows the parameter $e$ versus the Farey index $n$ 
for the periodic windows. 
Many of resonances $m_1:m_2$ corresponding to Farey sequence $F_n$ 
are confirmed, although some of them are not observed because of 
the possible lack of stability.

\begin{table}[t]
\caption{Farey sequence $F_n$ up to $n=14$.}

\begin{xy}
(0,20) *{ F_1 }="F_1", (10,20) *{ 0:1 }="0:1", (110,20) *{ 1:1 }="1:1",
(0,15) *{ F_2 }="F_2", (60,15) *{ 1:2 }="1:2",
(0,10) *{ F_3 }="F_3", (35,10) *{ 1:3 }="1:3", (85,10) *{ 2:3 }="2:3",
(0,5) *{ F_4 }="F_4", (22,5) *{ 1:4 }="1:4", (47,5) *{ 2:5 }="2:5",
                      (72,5) *{ 3:5 }="3:5", (97,5) *+[F]{ 3:4 }="3:4",
(0,0) *{ F_5 }="F_5", (16,0) *{ 1:5 }="1:5", (28,0) *{ 2:7 }="2:7",
                      (41,0) *{ 3:8 }="3:8", (53,0) *{ 3:7 }="3:7",
                      (66,0) *{ 4:7 }="4:7", (78,0) *{ 5:8 }="5:8",
                      (88,0) *+[F]{ 5:7 }="5:7", (103,0) *{ 4:5 }="4:5",
\ar @{-} "0:1";"1:2"
\ar @{-} "1:1";"1:2"
\ar @{-} "0:1";"1:3"
\ar @{-} "1:1";"2:3"
\ar @{-} "1:2";"1:3"
\ar @{-} "1:2";"2:3"
\ar @{-} "0:1";"1:4"
\ar @{-} "1:3";"1:4"
\ar @{-} "1:3";"2:5"
\ar @{-} "1:2";"2:5"
\ar @{-} "1:2";"3:5"
\ar @{-} "2:3";"3:5"
\ar @{-} "2:3";"3:4"
\ar @{-} "1:1";"3:4"
\ar @{-} "0:1";"1:5"
\ar @{-} "1:4";"1:5"
\ar @{-} "1:4";"2:7"
\ar @{-} "1:3";"2:7"
\ar @{-} "1:3";"3:8"
\ar @{-} "2:5";"3:8"
\ar @{-} "2:5";"3:7"
\ar @{-} "1:2";"3:7"
\ar @{-} "1:2";"4:7"
\ar @{-} "3:5";"4:7"
\ar @{-} "3:5";"5:8"
\ar @{-} "2:3";"5:8"
\ar @{-} "2:3";"5:7"
\ar @{-} "3:4";"5:7"
\ar @{-} "3:4";"4:5"
\ar @{-} "1:1";"4:5"

\end{xy}

\

\begin{xy}
(0,25) *{ F_4 }="F_4",                         (110,25) *+[F]{ 3:4 }="3:4",
(0,20) *{ F_5 }="F_5", (10,20) *+[F]{ 5:7 }="5:7", 
(0,15) *{ F_6 }="F_6", (60,15) *{ 8:11 }="8:11",
(0,10) *{ F_7 }="F_7", (35,10) *{ 13:18 }="13:18", (85,10) *{ 11:15 }="11:15",
(0,5) *{ F_8 }="F_8", (22,5) *{ 18:25 }="18:25", (47,5) *{ 21:29 }="21:29",
                      (72,5) *{ 19:26 }="19:26", (97,5) *{ 14:19 }="14:19",
(0,0) *{ F_9 }="F_9", (16,0) *{ 23\!:\!32 }="23:32", 
                      (28,0) *{ 31\!:\!43 }="31:43",
                      (41,0) *{ 34\!:\!47 }="34:47",
                      (53,0) *{ 29\!:\!40 }="29:40",
                      (66,0) *{ 27\!:\!37 }="27:37",
                      (78,0) *{ 30\!:\!41 }="30:41",
                      (91,0) *{ 25\!:\!34 }="25:34",
                      (103,0) *+[F]{ 17\!:\!23 }="17:23",
\ar @{-} "5:7";"8:11"
\ar @{-} "3:4";"8:11"
\ar @{-} "5:7";"13:18"
\ar @{-} "8:11";"13:18"
\ar @{-} "8:11";"11:15"
\ar @{-} "3:4";"11:15"
\ar @{-} "5:7";"18:25"
\ar @{-} "13:18";"18:25"
\ar @{-} "13:18";"21:29"
\ar @{-} "8:11";"21:29"
\ar @{-} "8:11";"19:26"
\ar @{-} "11:15";"19:26"
\ar @{-} "11:15";"14:19"
\ar @{-} "3:4";"14:19"
\ar @{-} "5:7";"23:32"
\ar @{-} "18:25";"23:32"
\ar @{-} "18:25";"31:43"
\ar @{-} "13:18";"31:43"
\ar @{-} "13:18";"34:47"
\ar @{-} "21:29";"34:47"
\ar @{-} "21:29";"29:40"
\ar @{-} "8:11";"29:40"
\ar @{-} "8:11";"27:37"
\ar @{-} "19:26";"27:37"
\ar @{-} "19:26";"30:41"
\ar @{-} "11:15";"30:41"
\ar @{-} "11:15";"25:34"
\ar @{-} "14:19";"25:34"
\ar @{-} "14:19";"17:23"
\ar @{-} "3:4";"17:23"

\end{xy}

\

\begin{xy}
(0,39) *{ F_4 }="F_4",                         (140,39) *+[F]{ 3:4 }="3:4",
(0,34) *{ F_9 }="F_9", (10,34) *+[F]{ 17:23 }="17:23", 
(0,29) *{ F_{10} }="F_10", (75,29) *{ 20:27 }="20:27",
(0,24) *{ F_{11} }="F_11", (42,24) *{ 37:50 }="37:50", (107,24) *{ 23:31 }="23:31",
(0,19) *{ F_{12} }="F_12", (26,19) *{ 54:73 }="54:73", (58,19) *{ 57:77 }="57:77",
                      (91,19) *{ 43:58 }="43:58", (123,19) *{ 26:35 }="26:35",
(0,14) *{ F_{13} }="F_13", (18,14) *{ 71:96 }="71:96", 
                          (34,14) *{ 91:123 }="91:123",
                          (50,14) *{ 94:127 }="94:127",
                          (66,14) *{ 77:104 }="77:104",
                          (83,14) *{ 63:85 }="63:85",
                          (99,14) *{ 66:89 }="66:89",
                          (115,14) *{ 49:66 }="49:66",
                          (131,14) *{ 29:39 }="29:39",
(0,7) *{ F_{14} }="F_14", (14,7) *{ 88\!:\!119 }="88:119", 
                          (30,7) *{ 145\!:\!196 }="145:196", 
                          (46,7) *{ 131\!:\!177 }="131:177", 
                          (62,7) *{ 134\!:\!181 }="134:181", 
                          (79,7) *{ 83\!:\!112 }="83:112", 
                          (95,7) *{ 109\!:\!147 }="109:147", 
                          (111,7) *{ 72\!:\!97 }="72:97", 
                          (127,7) *{ 55\!:\!74 }="55:74", 
(0,0) *{F_{14}},          (22,0) *{ 125\!:\!169 }="125:169", 
                          (38,0) *{ 128\!:\!173 }="128:173", 
                          (54,0) *{ 151\!:\!204 }="151:204", 
                          (70,0) *{ 97\!:\!131 }="97:131", 
                          (87,0) *{ 106\!:\!143 }="106:143", 
                          (103,0) *{ 89\!:\!120 }="89:120", 
                          (119,0) *{ 75\!:\!101 }="75:101", 
                          (135,0) *{ 32\!:\!43 }="32:43",


\ar @{-} "17:23";"20:27"
\ar @{-} "3:4";"20:27"
\ar @{-} "17:23";"37:50"
\ar @{-} "20:27";"37:50"
\ar @{-} "20:27";"23:31"
\ar @{-} "3:4";"23:31"
\ar @{-} "17:23";"54:73"
\ar @{-} "37:50";"54:73"
\ar @{-} "37:50";"57:77"
\ar @{-} "20:27";"57:77"
\ar @{-} "20:27";"43:58"
\ar @{-} "23:31";"43:58"
\ar @{-} "23:31";"26:35"
\ar @{-} "3:4";"26:35"
\ar @{-} "17:23";"71:96"
\ar @{-} "54:73";"71:96"
\ar @{-} "54:73";"91:123"
\ar @{-} "37:50";"91:123"
\ar @{-} "37:50";"94:127"
\ar @{-} "57:77";"94:127"
\ar @{-} "57:77";"77:104"
\ar @{-} "20:27";"77:104"
\ar @{-} "20:27";"63:85"
\ar @{-} "43:58";"63:85"
\ar @{-} "43:58";"66:89"
\ar @{-} "23:31";"66:89"
\ar @{-} "23:31";"49:66"
\ar @{-} "26:35";"49:66"
\ar @{-} "26:35";"29:39"
\ar @{-} "3:4";"29:39"
\ar @{-} "17:23";"88:119"
\ar @{-} "71:96";"88:119"
\ar @{-} "71:96";"125:169"
\ar @{-} "54:73";"125:169"
\ar @{-} "54:73";"145:196"
\ar @{-} "91:123";"145:196"
\ar @{-} "91:123";"128:173"
\ar @{-} "37:50";"128:173"
\ar @{-} "37:50";"131:177"
\ar @{-} "94:127";"131:177"
\ar @{-} "94:127";"151:204"
\ar @{-} "57:77";"151:204"
\ar @{-} "57:77";"134:181"
\ar @{-} "77:104";"134:181"
\ar @{-} "77:104";"97:131"
\ar @{-} "20:27";"97:131"
\ar @{-} "20:27";"83:112"
\ar @{-} "63:85";"83:112"
\ar @{-} "63:85";"106:143"
\ar @{-} "43:58";"106:143"
\ar @{-} "43:58";"109:147"
\ar @{-} "66:89";"109:147"
\ar @{-} "66:89";"89:120"
\ar @{-} "23:31";"89:120"
\ar @{-} "23:31";"72:97"
\ar @{-} "49:66";"72:97"
\ar @{-} "49:66";"75:101"
\ar @{-} "26:35";"75:101"
\ar @{-} "26:35";"55:74"
\ar @{-} "29:39";"55:74"
\ar @{-} "29:39";"32:43"
\ar @{-} "3:4";"32:43"
\end{xy}

\end{table}

\begin{table}[t]
\caption{Observed periodic windows. $m_1:m_2$ denotes the resonance condition, 
$n$ the Farey index, and $e_l$ $(e_u)$ the lower (upper) limit of $e$, 
respectively. NF denotes that the resonant periodic 
solutions are not found. 
\
}
\begin{center}
\begin{tabular}{|c|c|c|c|}
\hline
$m_1:m_2$ & $n$ &  $e_l$ & $e_u$  \\
\hline
17:23 & 9 & 0.02028 & 0.02112  \\
\hline
88:119 & 14 & 0.021625  & 0.021640  \\
\hline
71:96 & 13 &  0.021794 & 0.021817  \\
\hline
125:169 & 14 & 0.021924  & 0.021926  \\
\hline
54:73 & 12 & 0.022056 & 0.022113  \\
\hline
145:196 & 14  & 0.022217 & 0.022219  \\
\hline
91:123 & 13 & 0.022287  & 0.022299 \\
\hline
128:173 & 14  & 0.022373 & 0.022377 \\
\hline
37:50 & 11 & 0.022513 & 0.02269 \\
\hline
131:177 & 14 & 0.022813  & 0.022822  \\
\hline
94:127 & 13  & 0.022873 & 0.022898  \\
\hline
151:204 & 14  & 0.022937 & 0.022947 \\
\hline
57:77 & 12 & 0.023013 & 0.023087 \\
\hline
134:181 & 14  &0.023147 & 0.023162 \\
\hline
77:104 & 13 & 0.02320 & 0.023251 \\
\hline
97:131 & 14 & 0.023301 & 0.023332 \\
\hline
20:27 & 10  & 0.02350 & 0.02446   \\
\hline
83:112 & 14  & 0.024650 & 0.024656   \\
\hline
83:112 & 14  & 0.02489 & 0.0249356   \\
\hline
63:85 & 13  & 0.024594 & 0.024605  \\
\hline
106:143 & 14  & 0.024698  & 0.0247   \\
\hline
43:58 & 12  & 0.02476 & 0.02479 \\
\hline
43:58 & 12  & 0.025015 & 0.025085 \\
\hline
109:147 & 14  & NF & NF \\
\hline
66:89 & 13  & NF & NF  \\
\hline
89:120 & 14  &0.025098 &0.025102 \\
\hline
23:31 & 11 & 0.02521 & 0.0255  \\
\hline
23:31 & 11 & 0.0259 & 0.02613  \\
\hline
72:97 & 14 & NF & NF  \\
\hline
49:66 & 13 & NF & NF  \\
\hline
75:101 & 14 & 0.0261475 & 0.026148 \\
\hline
26:35 & 12 & 0.02626 & 0.02636 \\
\hline
26:35 & 12 & 0.02705 & 0.02711 \\
\hline
55:74 & 14 & NF & NF \\
\hline
29:39 & 13 & 0.027675 &0.02772   \\
\hline
32:43 & 14 &0.0274  & 0.02743  \\
\hline
3:4 & 4 & 0.02935 & 0.057  \\
\hline

\end{tabular}
\end{center}
\end{table}

\section{Conclusions}

In conclusion, bifurcation structures similar to \cite{Bekki1,Bekki2,Umeki1} 
is observed in the Langford equation; 
the coexistence of the double periodicity, frequency-locking and 
period-doubling bifurcations. 
The Langford equation would be one of the most illustrative ODE 
systems having double periodicity since it consists of only three 
variables. 
Each periodic attractor represents a corresponding knot 
in three dimensional space, 
as discussed in \cite{Kauffman} for the Lorenz system and in \cite{Bekki2} for 
the magnetoconvection system. 

Recent progress of cost performance in computer hardware 
would make it possible to judge whether similar double periodicity 
and frequency-locking appear in the high dimensional Navier-Stokes 
turbulence and other chaotic systems. The method to judge quasi periodicity 
would be basically the same as that stated in the presented work. 

This study was motivated by comments by Professor Bekki at the meeting of 
Japan Physical Society several years ago. 
The author is grateful to Professor Yamagata for support of research 
and to Professor Langford for sending the author his reprints.

\newpage

\begin{figure}[H]
  \begin{center}
  \vspace{0mm}
    \begin{tabular}{c}
      \resizebox{70mm}{!}{\includegraphics[angle=0]{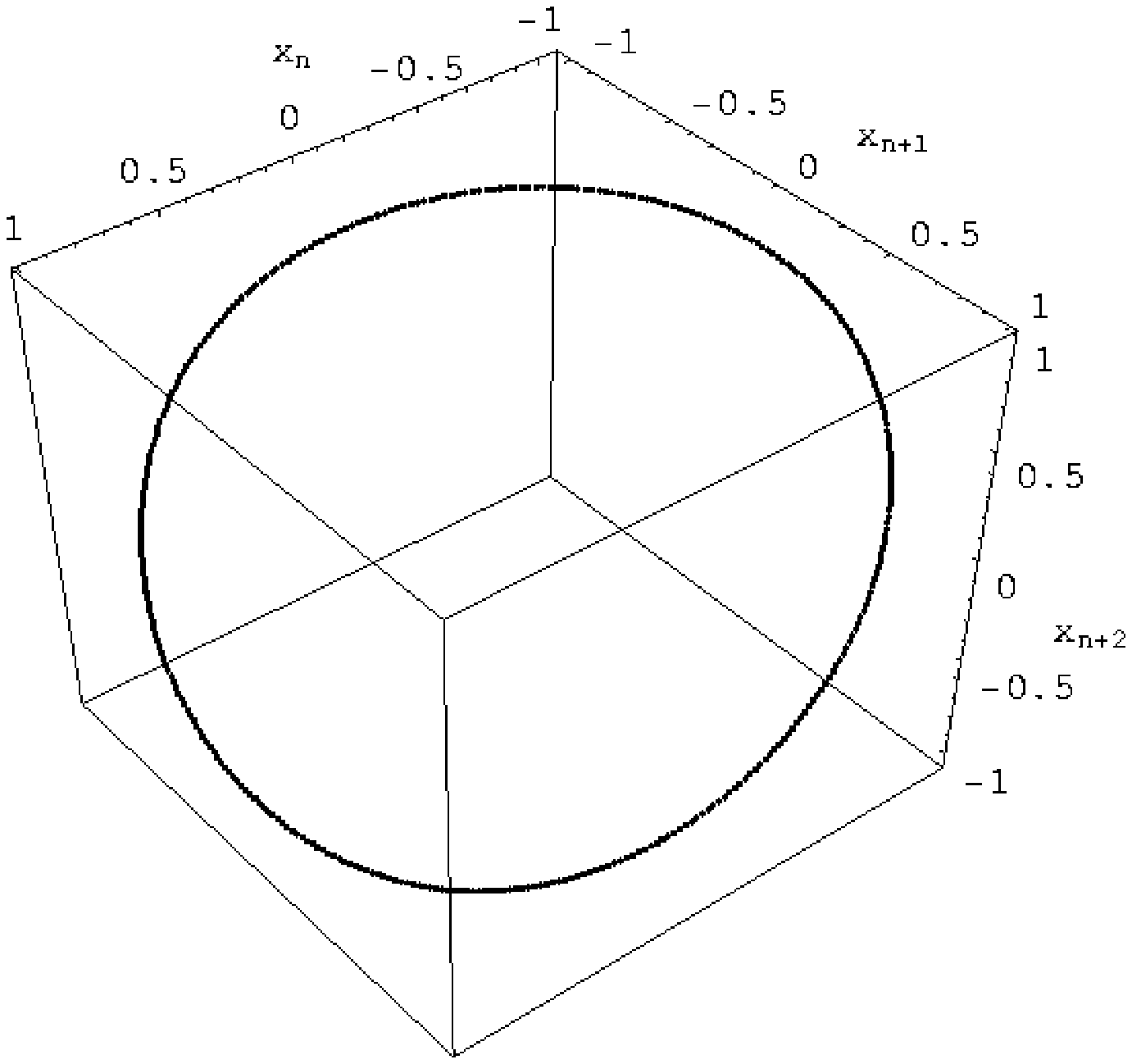}} \\
      \resizebox{70mm}{!}{\includegraphics[angle=0]{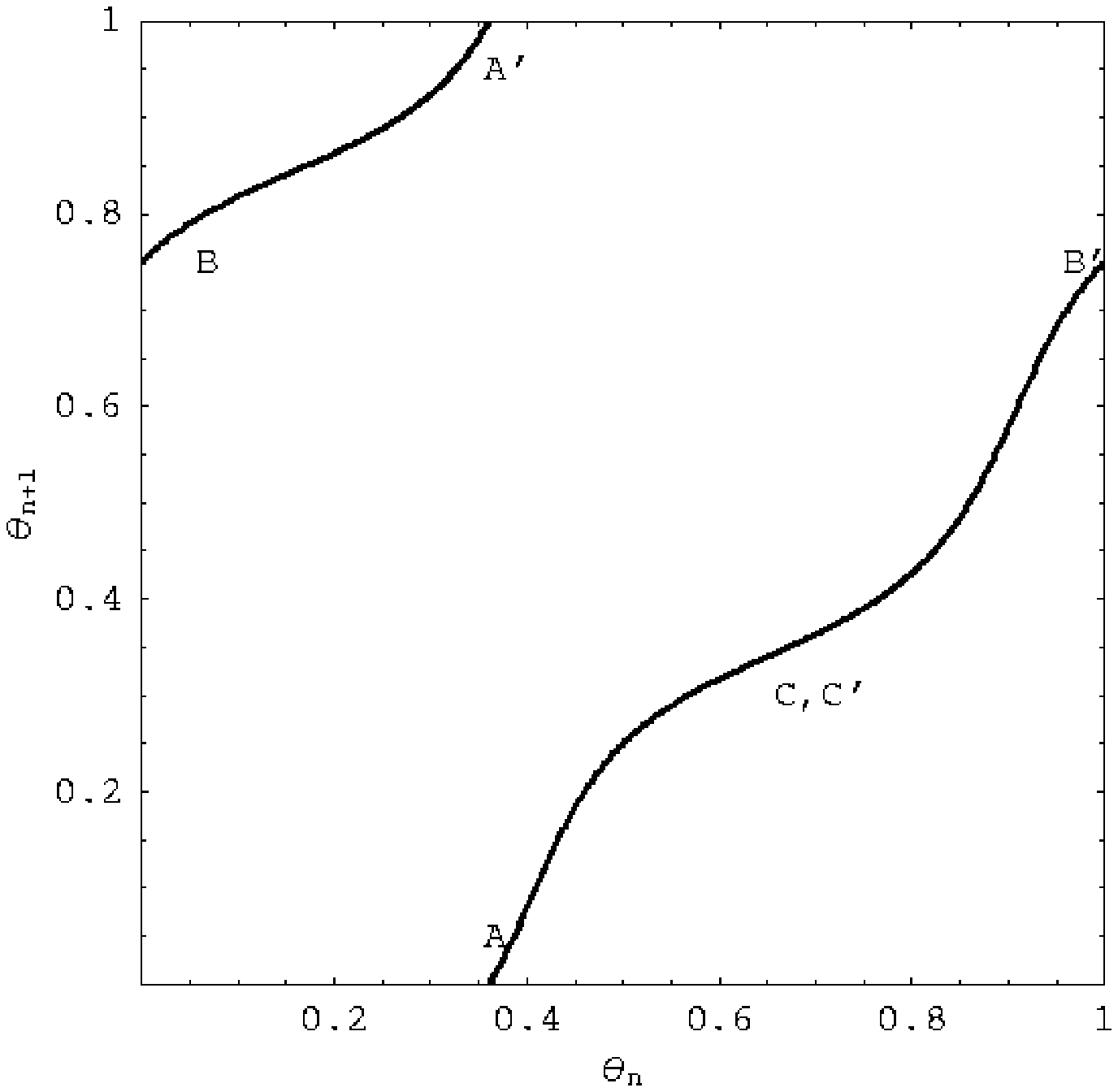}} \\
      \resizebox{70mm}{!}{\includegraphics[angle=0]{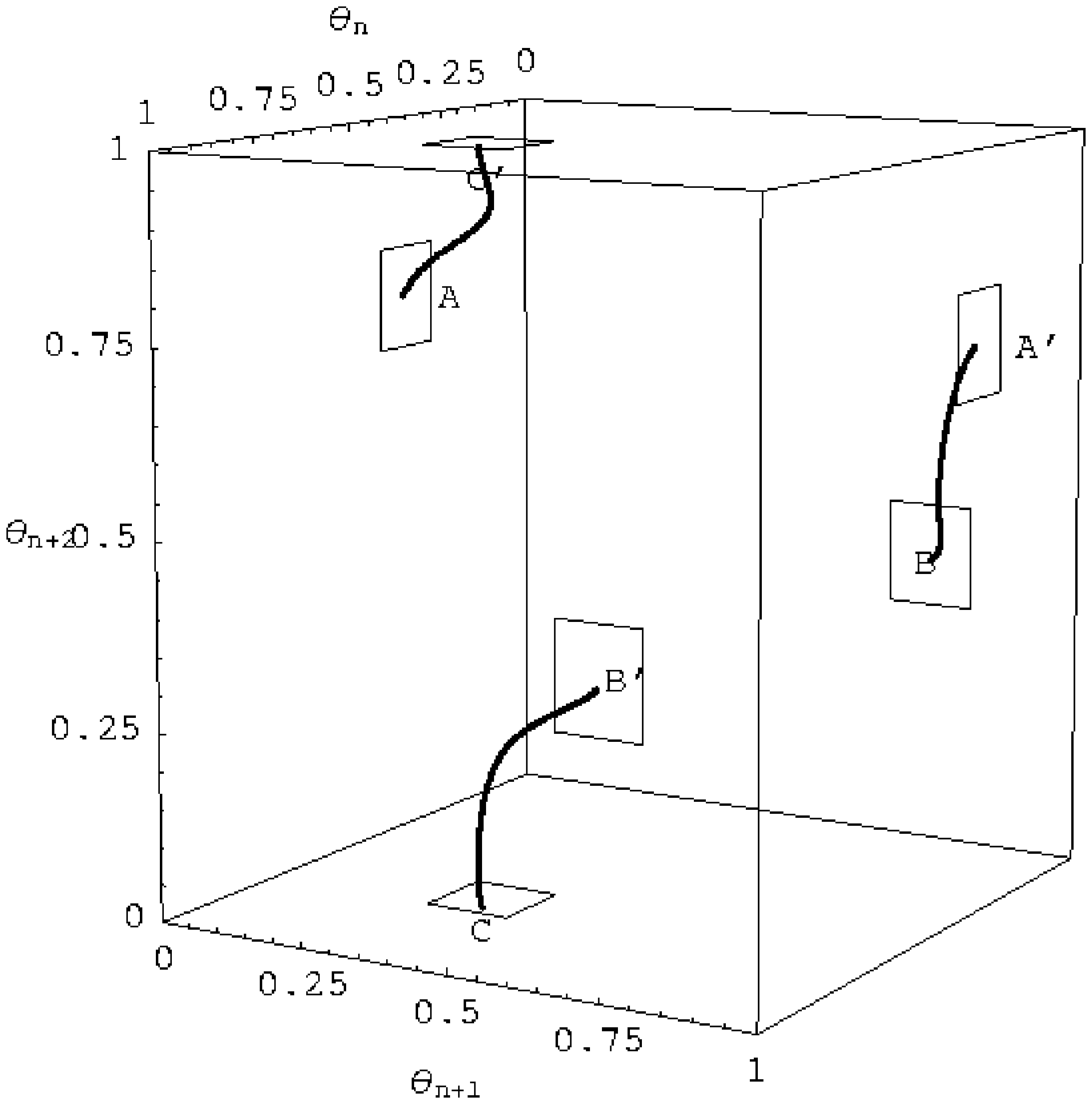}} \\
    \end{tabular}
    \caption{
Poincar\'e plot of (\ref{eq1}) for the double periodicity $a=0$ in (a) the $(x_n,x_{n+1},x_{n+2})$ space, 
on (b) the $(\theta_n, \theta_{n+1}))$ plane, and in 
(c) the $(\theta_n,\theta_{n+1},\theta_{n+2})$ space. 
Points A, B and A', B' in Figures 1(a) and 1(b), and C, C' in Figure 1(c), 
which are separated due to modulus $2\pi$ in the argument, are the same points. 
In order to see which surface the points lie on, the rectangles including 
the points are added in Figure 1(c).
}
    \label{fig1}
  \end{center}
\end{figure}

\begin{figure}[H]
  \begin{center}
  \vspace{0mm}
    \begin{tabular}{c}
      \resizebox{70mm}{!}{\includegraphics[angle=0]{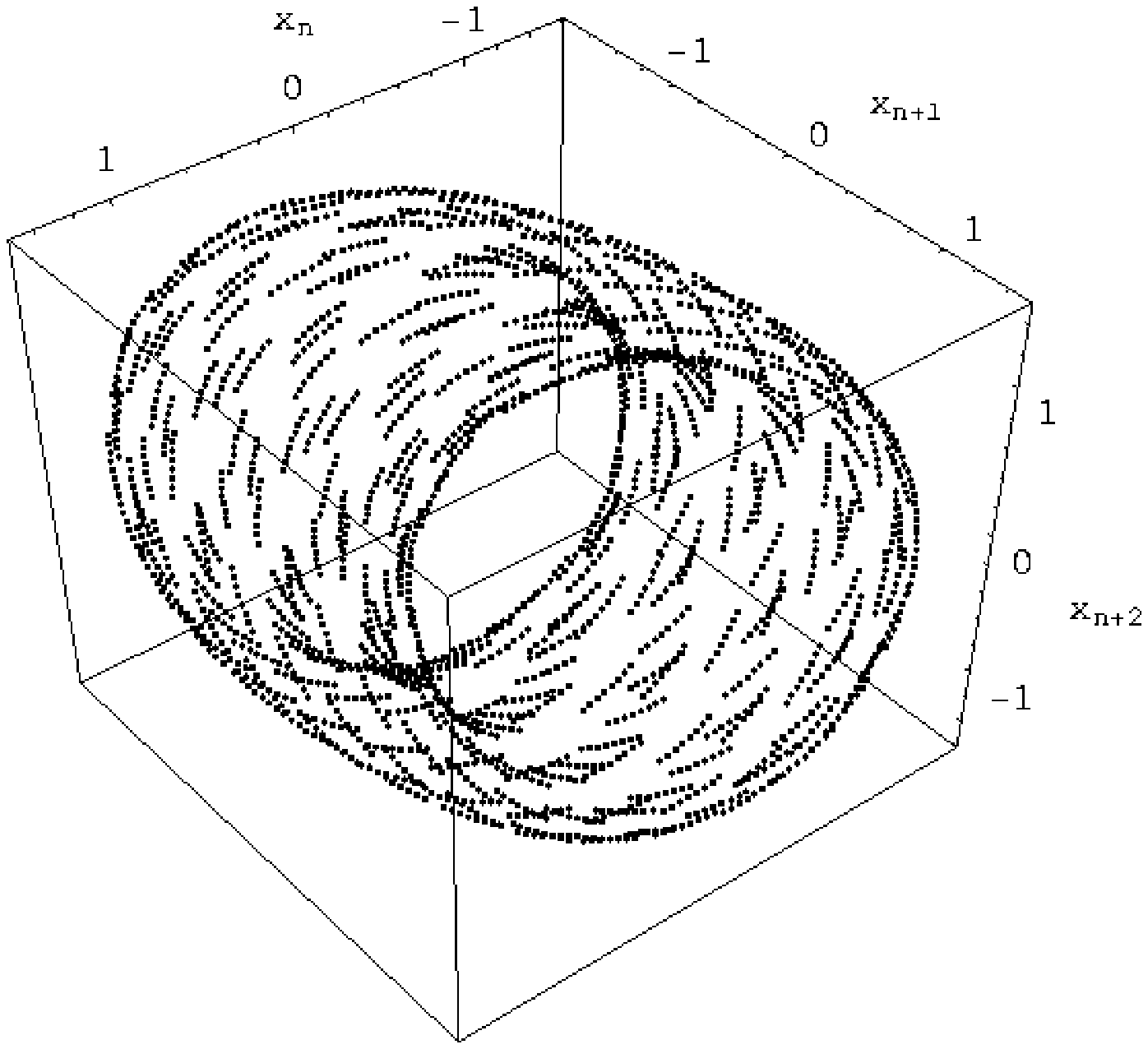}} \\
      \resizebox{70mm}{!}{\includegraphics[angle=0]{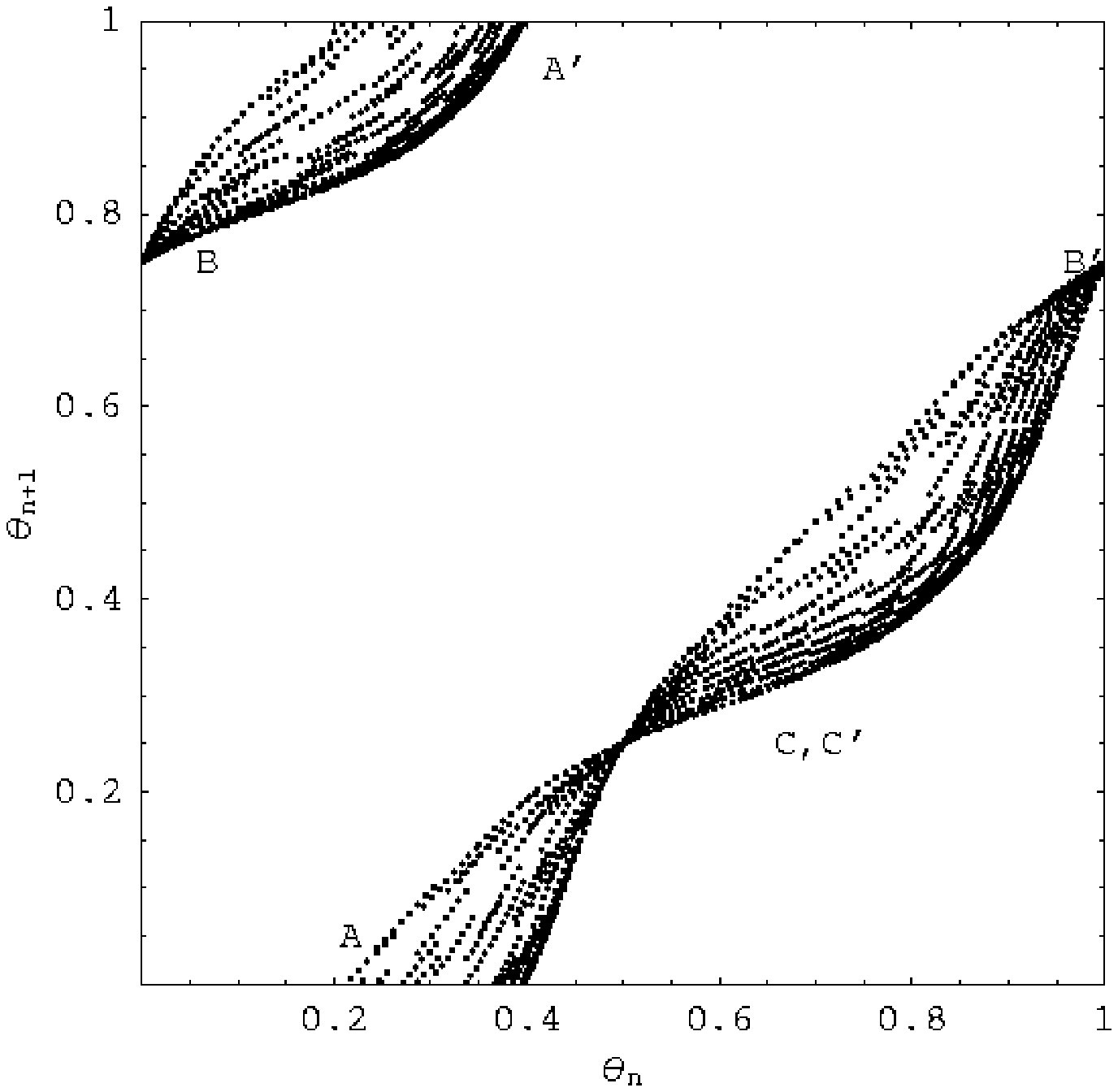}} \\
      \resizebox{70mm}{!}{\includegraphics[angle=0]{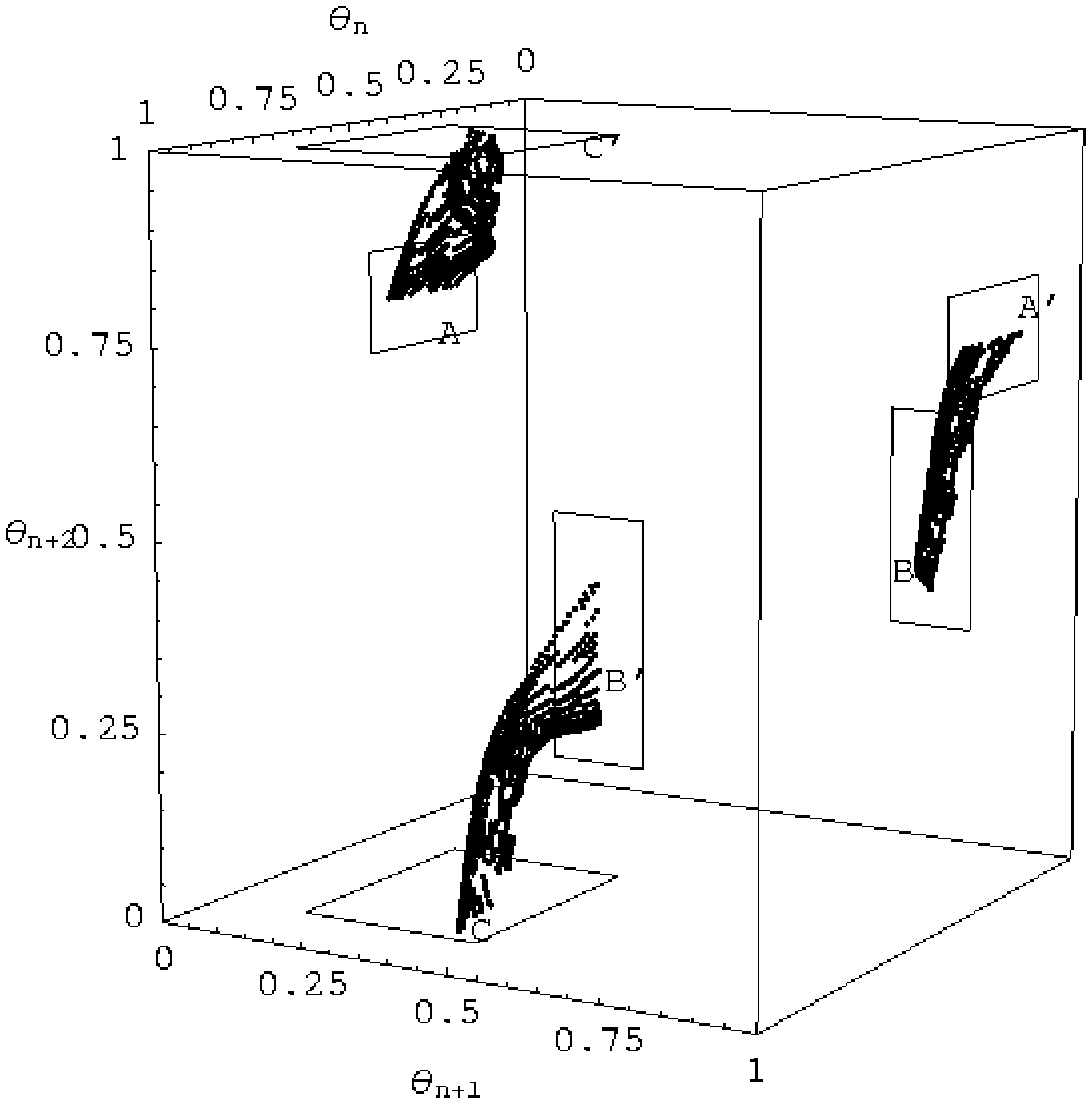}} \\
    \end{tabular}
    \caption{
Poincar\'e plot of (\ref{eq1}) for the triple periodicity $a=0.5,\omega=\sqrt{2}$ in 
(a) the $(x_n,x_{n+1},x_{n+2})$ space, 
on (b) the $(\theta_n, \theta_{n+1})$ plane, and in (c) 
the $(\theta_n,\theta_{n+1},\theta_{n+2})$ space. The points are on the 
surface in Figure 2(a) and 2(c). As $a$ in Eq. (\ref{eq1}) increases, the 
width of the surface becomes large.}
    \label{fig2}
  \end{center}
\end{figure}

\begin{figure}[H]
  \begin{center}
  \vspace{0mm}
    \begin{tabular}{c}
      \resizebox{105mm}{!}{\includegraphics[angle=0]{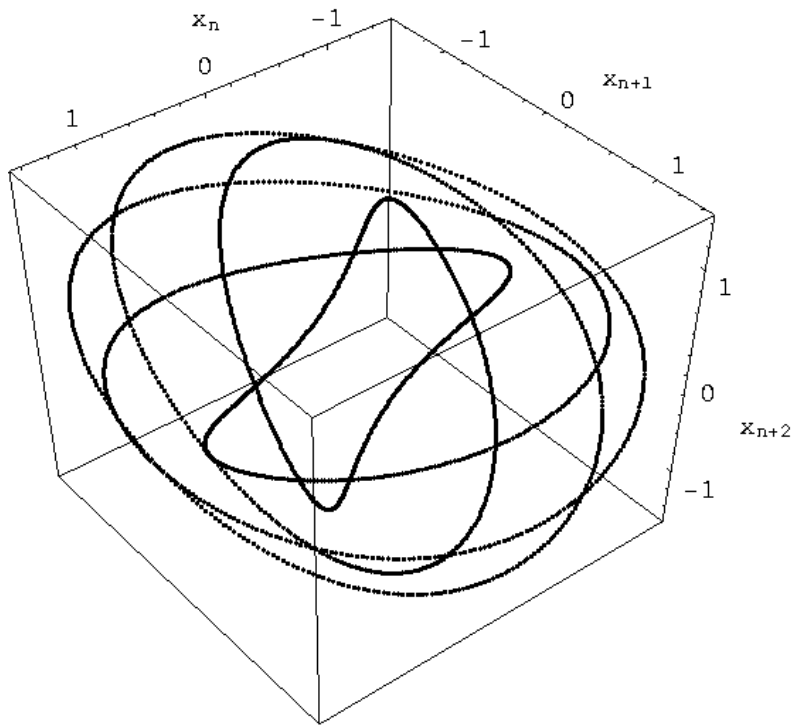}} \\
      \resizebox{105mm}{!}{\includegraphics[angle=0]{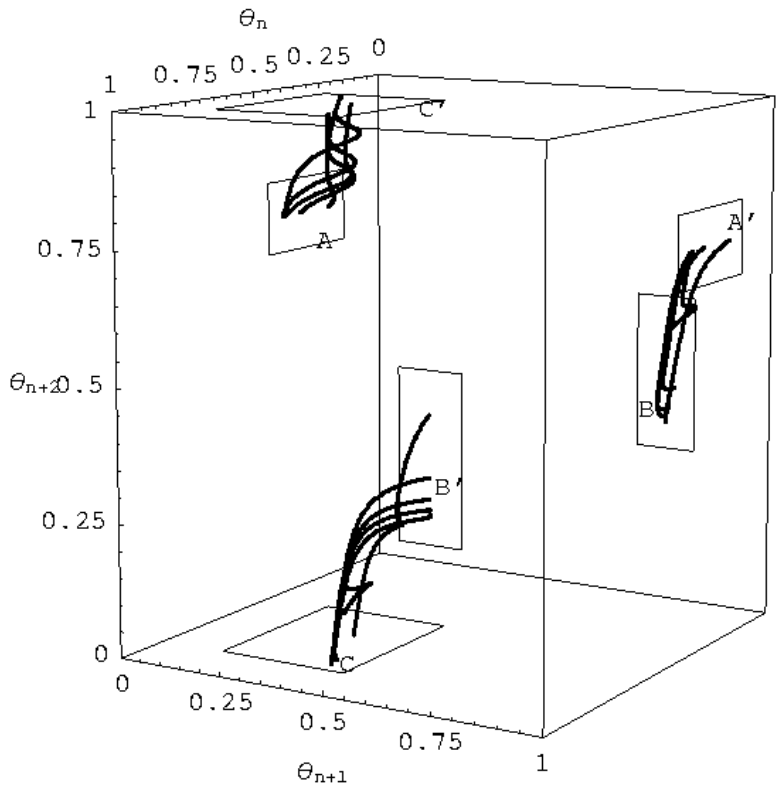}} \\
    \end{tabular}
    \caption{
Poincar\'e plot of (\ref{eq1}) for the frequency-locked double periodicity $a=0.5,\omega=1.4 $ in 
(a) the $(x_n,x_{n+1},x_{n+2})$ space and (b) the $(\theta_n,\theta_{n+1},\theta_{n+2})$ space. 
}
    \label{fig3}
  \end{center}
\end{figure}

\begin{figure}[H]
  \begin{center}
  \vspace{0mm}
    \begin{tabular}{c}
      \resizebox{136mm}{!}{\includegraphics[angle=90]{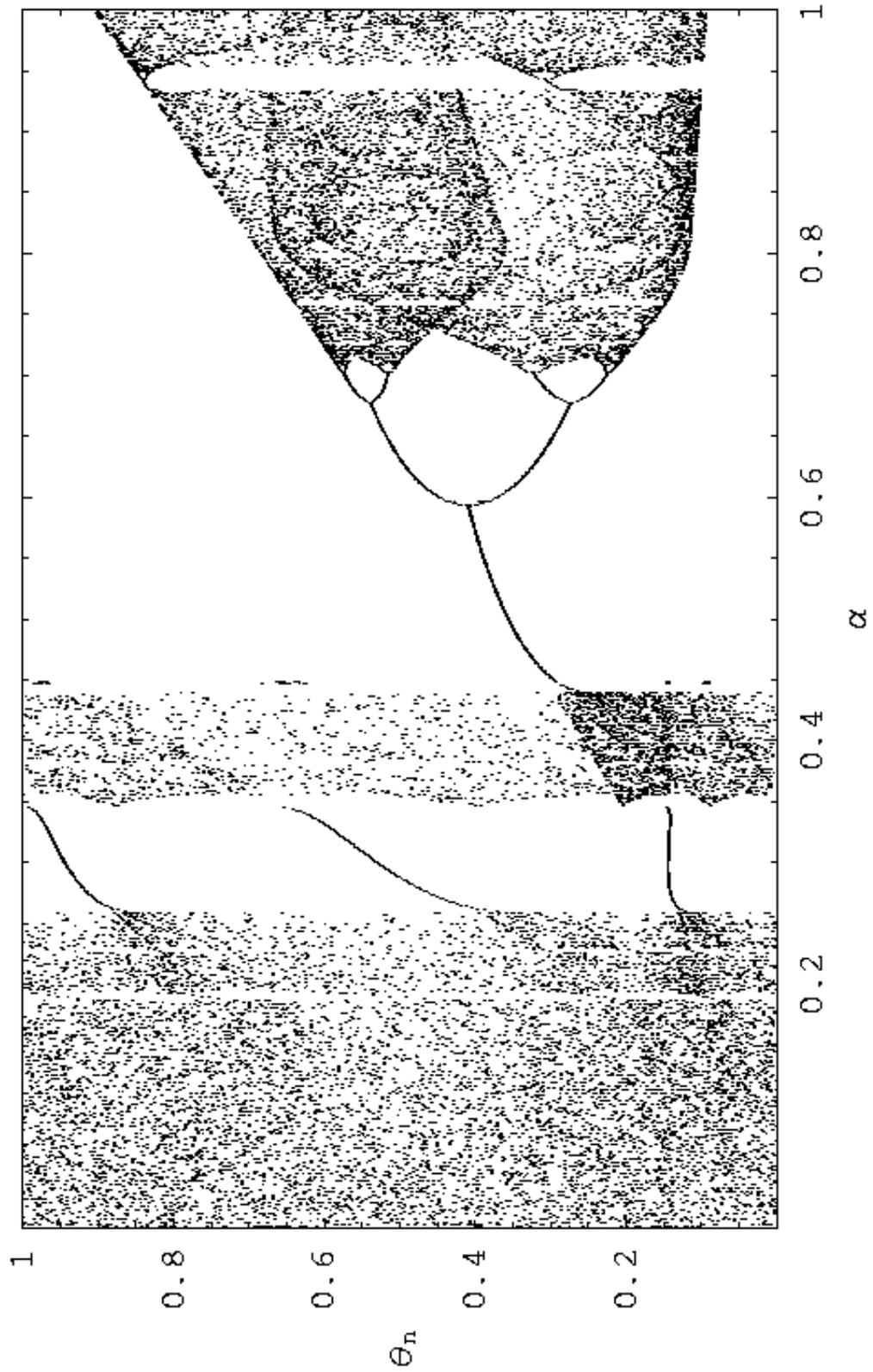}} \\
    \end{tabular}
    \caption{
Bifurcation diagram of the sine-circle map (\ref{eq8}). 
The parameters $\Omega$ and $K$ are related as (\ref{eq9}). 40000 points are randomly selected from 
the data for a clear view.}
    \label{fig4}
  \end{center}
\end{figure}

\begin{figure}[H]
  \begin{center}
  \vspace{0mm}
    \begin{tabular}{c}
      \resizebox{136mm}{!}{\includegraphics[angle=90]{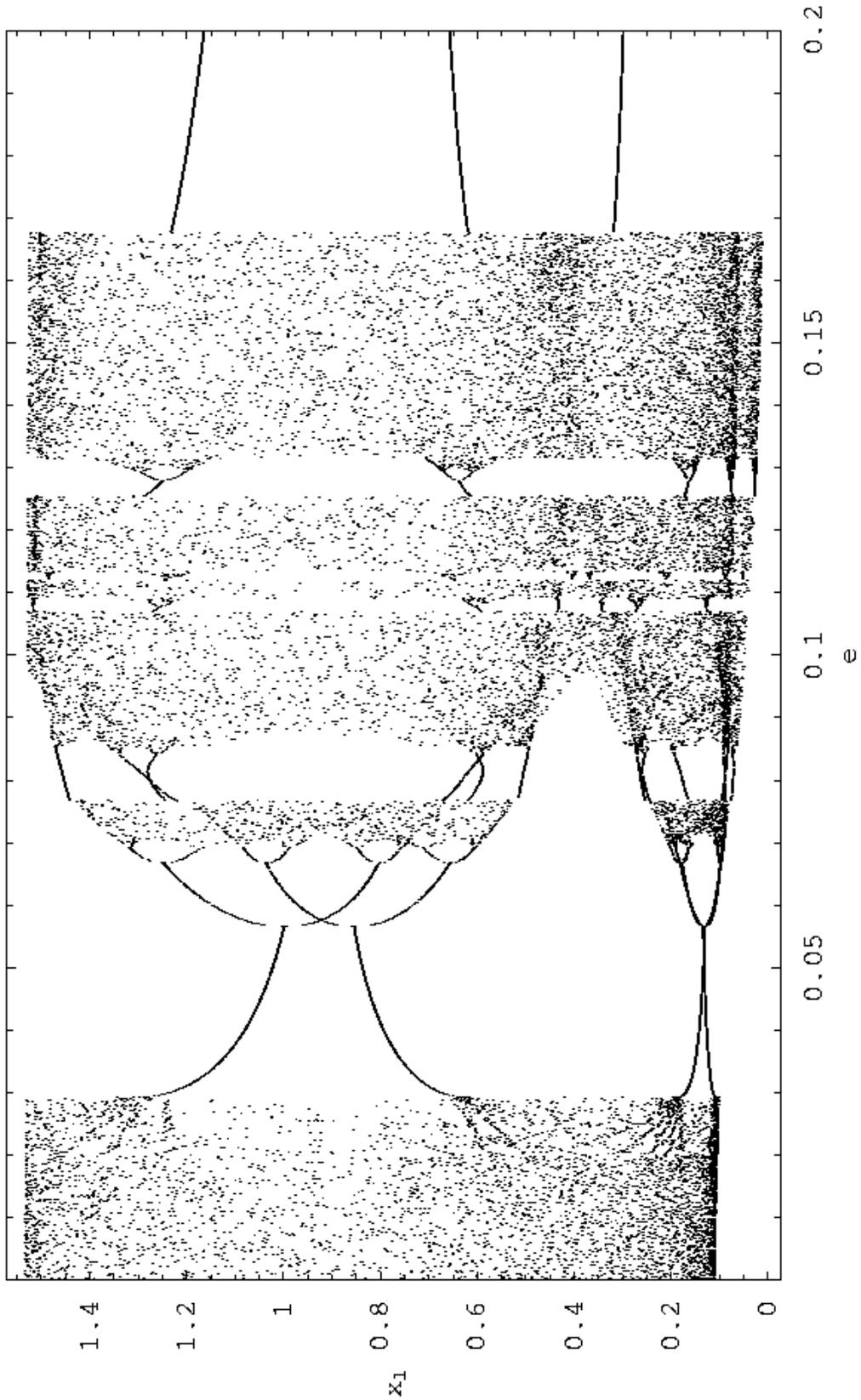}} \\
    \end{tabular}
    \caption{
Bifurcation diagram of the Langford equation for $0<e<0.2$. 40000 points are randomly selected from 
the data for a clear view. In the following figures, bifurcation diagrams are 
for the Langford equation.}
    \label{fig5}
  \end{center}
\end{figure}

\begin{figure}[H]
  \begin{center}
  \vspace{0mm}
    \begin{tabular}{c}
      \resizebox{136mm}{!}{\includegraphics[angle=90]{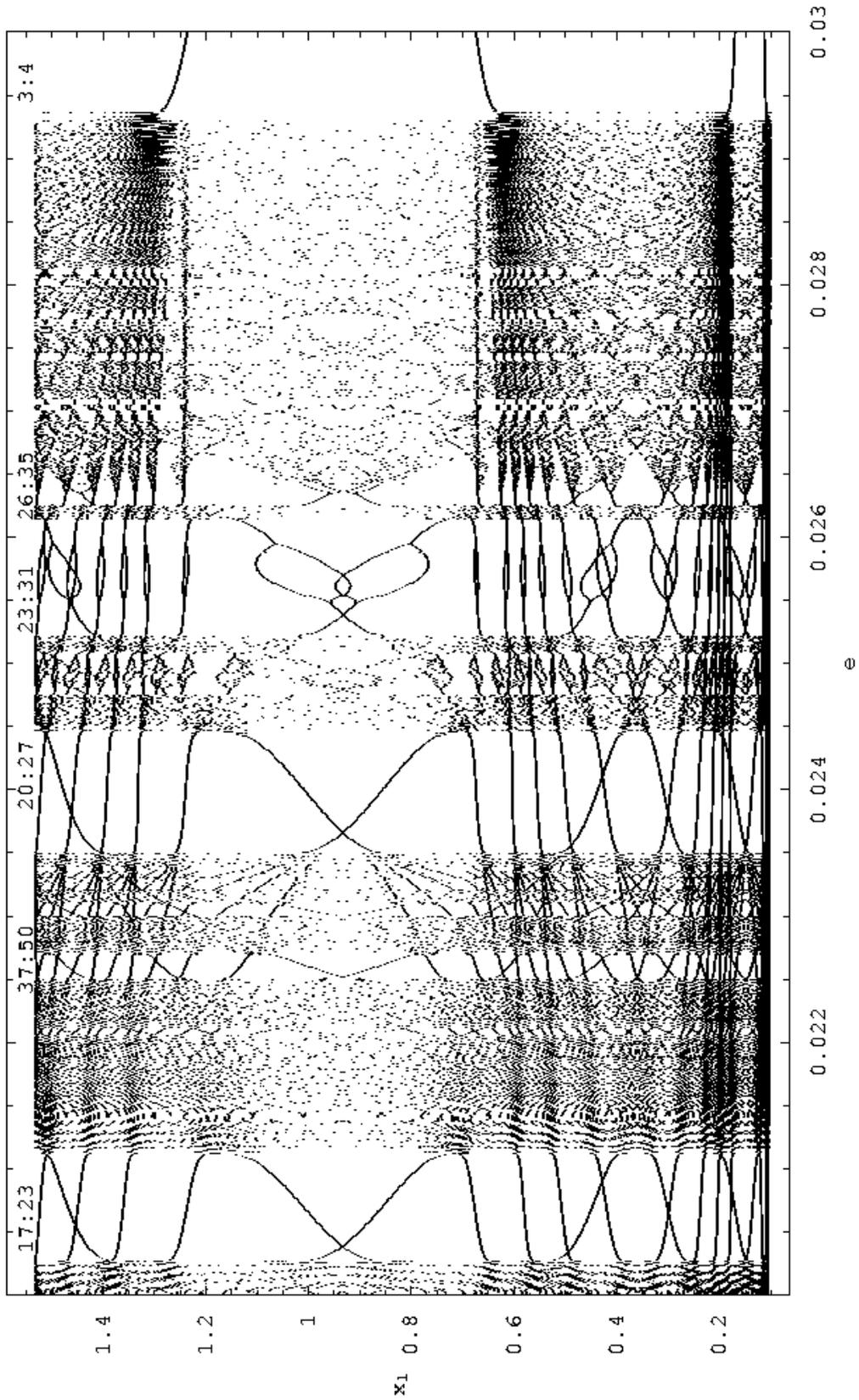}} \\
    \end{tabular}
    \caption{
Enlarged bifurcation diagram for $0.02<e<0.03$. 
}
    \label{fig6}
  \end{center}
\end{figure}

\begin{figure}[H]
  \begin{center}
  \vspace{0mm}
    \begin{tabular}{c}
      \resizebox{136mm}{!}{\includegraphics[angle=90]{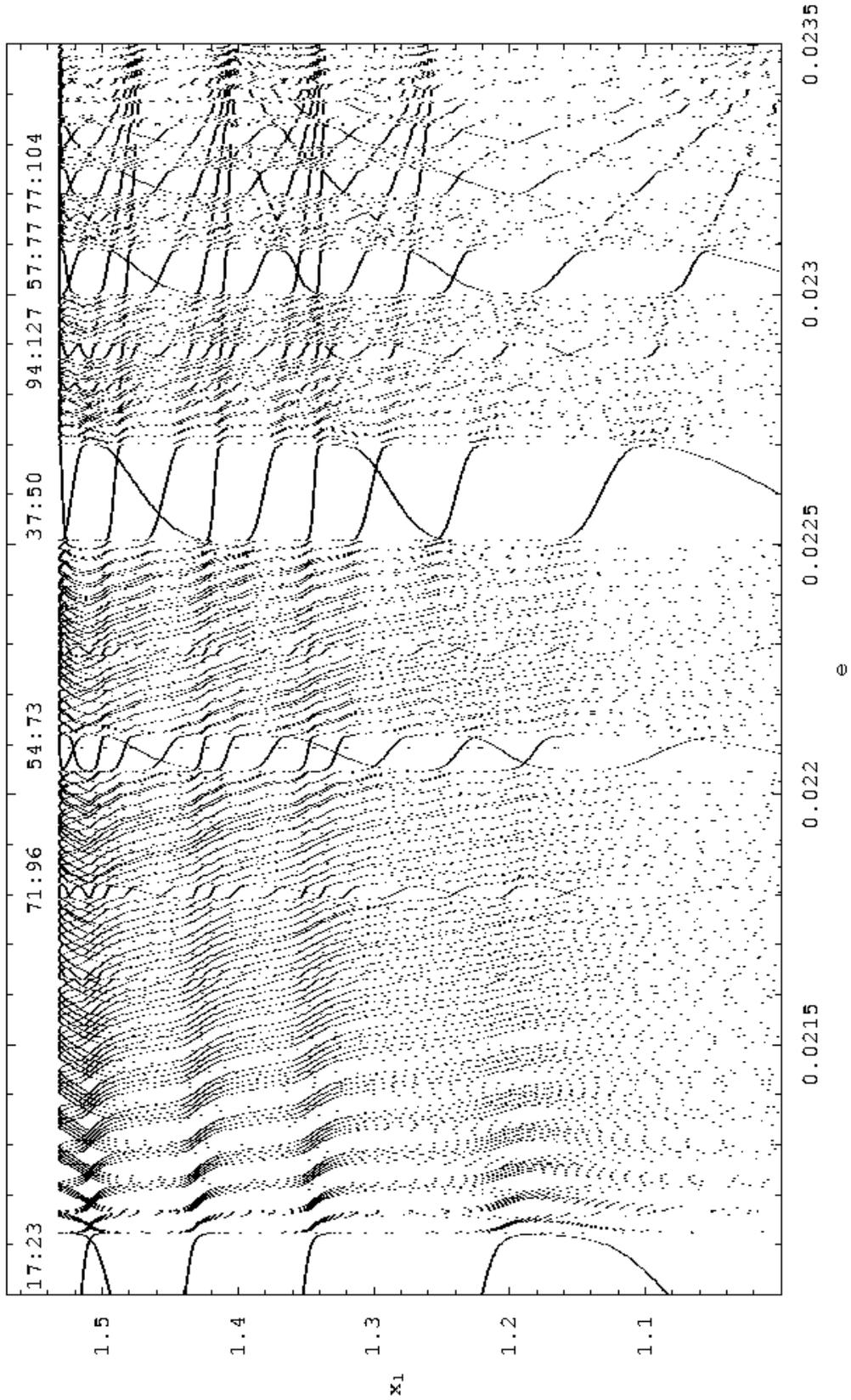}} \\
    \end{tabular}
    \caption{
Bifurcation diagram enlarged again for $0.021<e<0.0235$. 
}
    \label{fig7}
  \end{center}
\end{figure}

\begin{figure}[H]
  \begin{center}
  \vspace{0mm}
    \begin{tabular}{c}
      \resizebox{136mm}{!}{\includegraphics[angle=90]{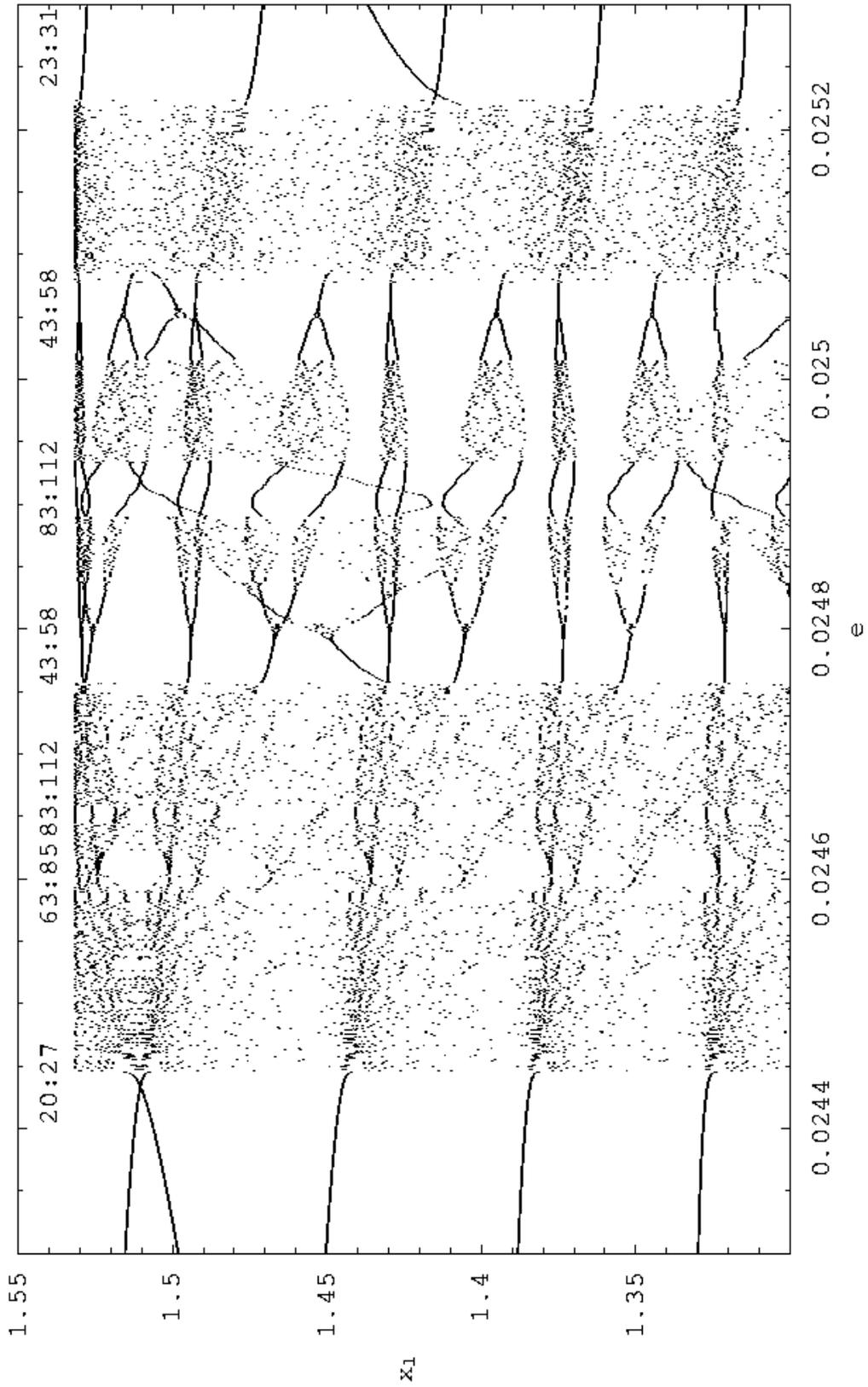}} \\
    \end{tabular}
    \caption{
Enlarged bifurcation diagram for $0.0243<e<0.0253$. 
Two parameter regions for 43:58 and 83:112 resonances are observed. 
The period-doubling bifurcation follows from 43:58 periodic solutions.  
}
    \label{fig8}
  \end{center}
\end{figure}

\begin{figure}[H]
  \begin{center}
  \vspace{0mm}
    \begin{tabular}{c}
      \resizebox{136mm}{!}{\includegraphics[angle=0]{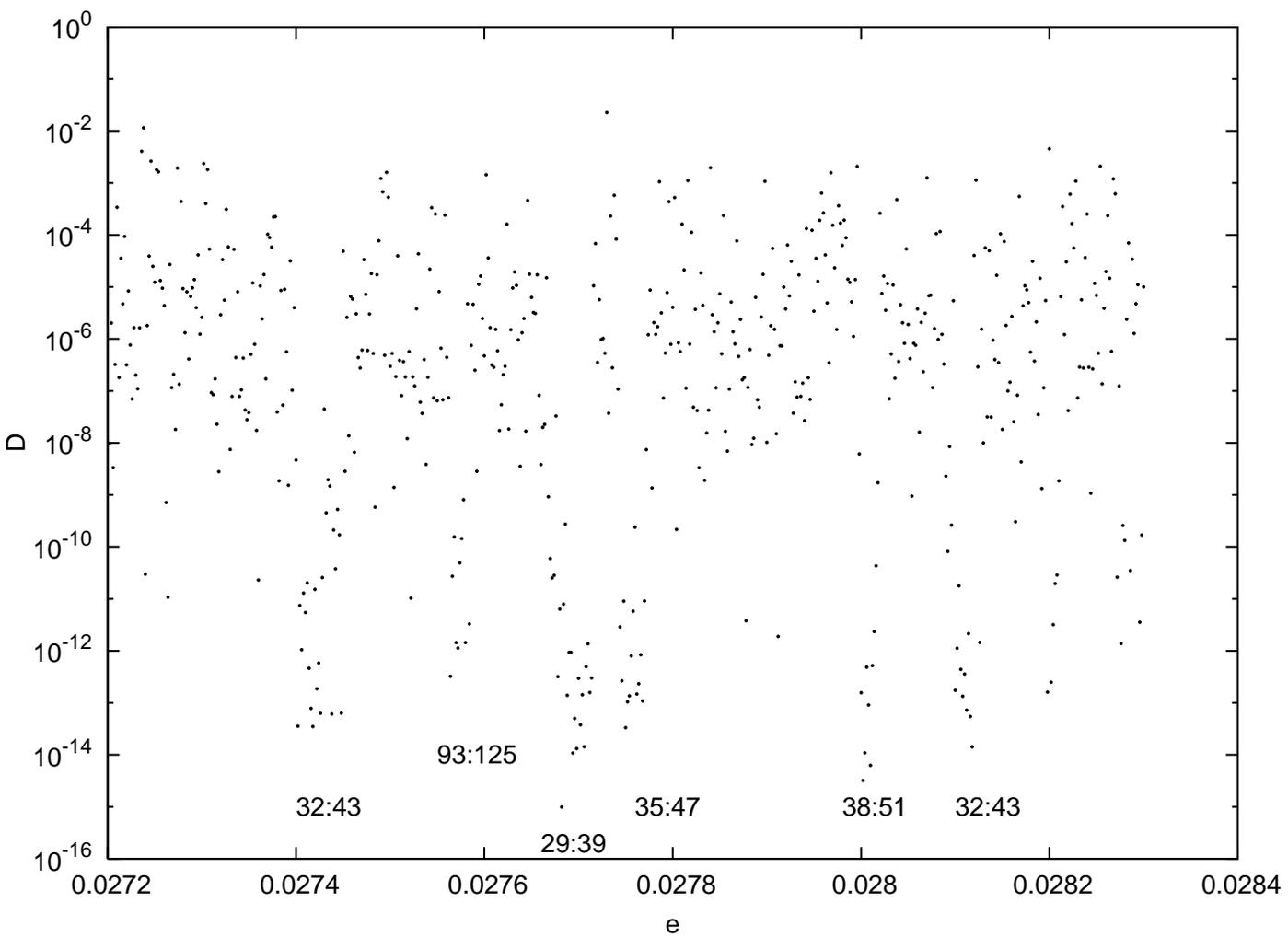}} \\
    \end{tabular}
    \caption{
The parameter $e$ versus the square of the minimum distance $D$ of numerical solutions showing its 
accuracy of recurrence for resonant periodicity. 
}
    \label{fig9}
  \end{center}
\end{figure}

\begin{figure}[H]
  \begin{center}
  \vspace{0mm}
    \begin{tabular}{c}
      \resizebox{136mm}{!}{\includegraphics[angle=0]{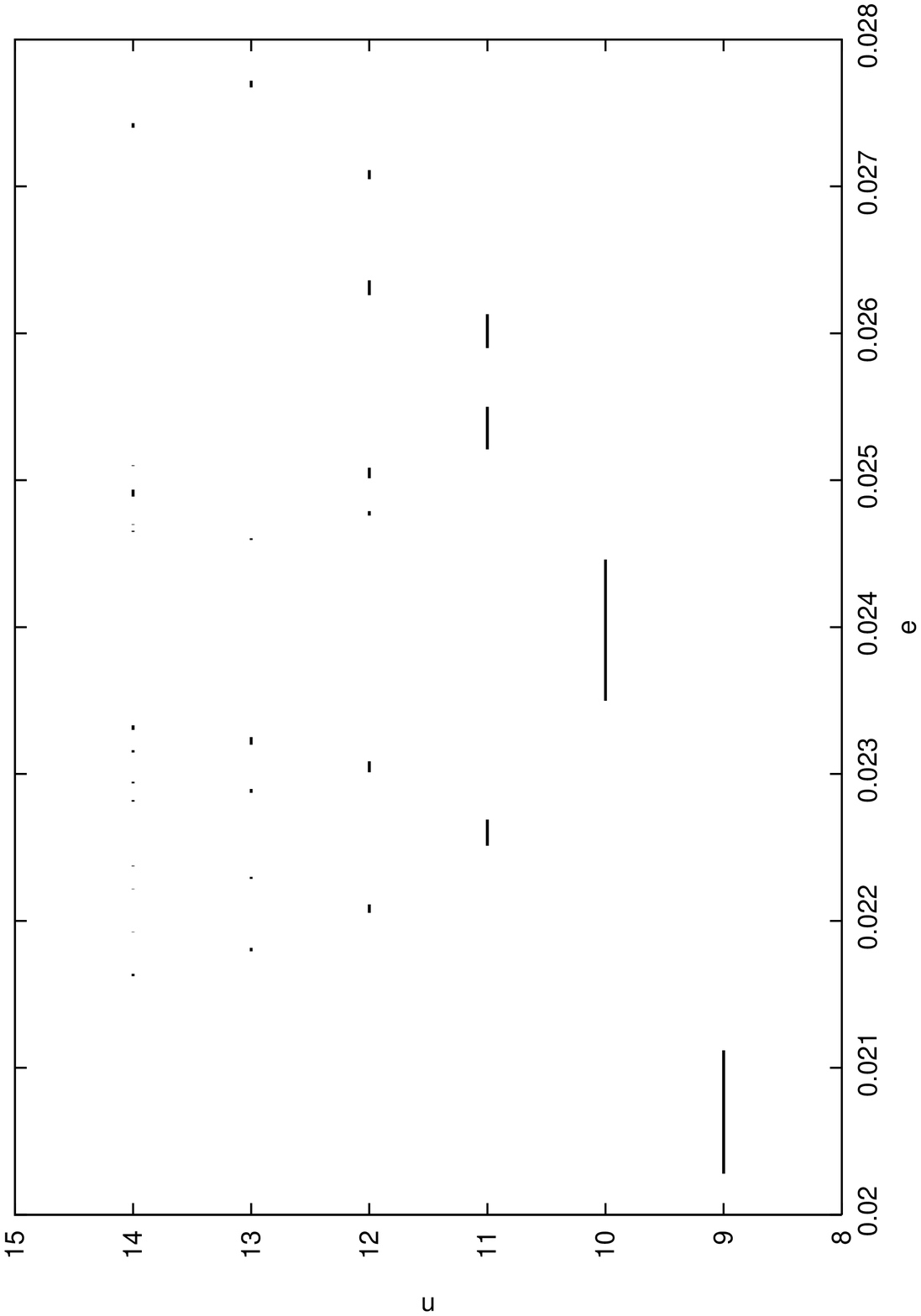}} \\
    \end{tabular}
    \caption{
The parameter $e$ versus the Farey index $n$ for periodic windows. 
}
    \label{fig10}
  \end{center}
\end{figure}


\begin{thebibliography}{99} 

\bibitem{Bekki1} N. Bekki and T. Karakisawa, 
Devil's Staircase in a Dissipative Fifth-Order System,
J. Phys. Soc. Jpn. \textbf{69} (2000) 2443. 
\bibitem{Bekki2} N. Bekki, Torus Knot in a Dissipative Fifth-Order System,
J. Phys. Soc. Jpn. \textbf{69} (2000) 295. 
\bibitem{Kauffman} L. H. Kauffman, {\it Knots and Physics,} Third Edition. World Scientific. 
(2001) 501. 
\bibitem{Langford1} W. F. Langford. Chaotic Dynamics in the Unfoldings 
of Degenerate Bifurcations, 
{\it Proceedings of the International Symposium on Applied Mathematics and Information Science,
Kyoto University, Japan, March 29-31} (1982)
\bibitem{Langford2} W. F. Langford. A Review of Interactions of Hopf 
and Steady-state Bifurcations,  
{\it Nonlinear Dynamics and Turbulence, 
edited by G. I. Barenblatt, G. Iooss, and D. D. Joseph} (1983) 215. 
\bibitem{Langford3} W. F. Langford. Numerical Studies of Torus Bifurcations,
{\it International Series of Numerical Mathematics} {\bf 70} (1984) 285.
\bibitem{Ruelle} D. Ruelle and F. Takens, On the Nature of Turbulence, 
Communications in Mathematical Physics, \textbf{20} (1971) 167. 
\bibitem{Umeki1} M. Umeki, 
Bifurcations and Chaos in a Six-dimensional Turbulence Model of Gledzer, 
J. Phys. Soc. Jpn. \textbf{76} (2007) 043401.
\end{thebibliography}
\end{document}